# Experimental observation of spatial jitters of a triple-pulse x-ray source based on the pinhole imaging technique

Yi Wang, Zhiyong Yang, Xiaobing Jing, Qin Li, Hengsong Ding, Zhiyong Dai

Key Laboratory of Pulsed Power, Institute of Fluid Physics, China Academy of Engineering Physics, Mianyang 621900, Sichuan Province, China

**Abstract:** In high-energy flash radiography, scattered photons will degrade the acquiring image, which limits the resolving power of the interface and density of the dense object. The application of large anti-scatter grid is capable of remarkably decreasing scattered photons, whereas requires a very stable source position in order to reduce the loss of signal photons in the grid structure. The pinhole imaging technique is applied to observe spatial jitters of a triple-pulse radiographic source produced by a linear induction accelerator. Numerical simulations are taken to analyze the performance of the imaging technique with same or close parameters of the pinhole object and experimental alignment Experiments are carried out to observe spatial jitters of the source between different measurements. Deviations of the source position between different pulses are also measured in each experiment.

## 1. Introduction

Flash radiography is an extraordinary diagnostic technique to investigate the hydrodynamic process of high explosives.[1] The radiographic source is generally produced by a linear induction accelerator (LIA), which accelerates the electron beam pulse to ~MeV energy and then focuses it onto a high-Z convertor target to generate x-ray photons through bremsstrahlung radiation.[2-6] The temporal width of the pulse is usually tens of nanoseconds, which enables a recording of an inner stopped-motion image of the dense object. In the experiment of flash radiography, photon scattering occurs due to the interaction with all objects placed in the light field, which is a major obstacle for acquiring fine details of the interface and the density of the object since the resolution of the obtained image is greatly reduced by scattered photons.[7,8]

During passing decades, a substantial number of efforts were devoted to the investigation of photon scatter properties and anti-scatter techniques. Various collimators made of heavy materials have been proposed to reduce scatter background and improve contrast in radiography.[9-11] However, the application of collimators placed between the object and the light source is usually accompanied with a loss of abaxial image information of the object. Since 1990s, the Los Alamos National Laboratory has developed a structure of large anti-scatter grid with very high grid ratios, which remarkably reduced scattered photons for radiography.[12] The scintillator array of the image receiving system is directly pinned to the grid, the correspondence of which is perfectly matched to be one-scintillator-pixel to one-grid-hole. Since the anti-scatter grid is designed to fit the directions in which radiation photons are emitted, it ought to be strictly aligned to focus exactly at the center of the light source. A spatial jitter of the source will result in a loss of primary radiation in the grid structure.[13] The higher the grid ratio is, the more sensitively the spatial jitter of the source will degrade to the image. In this paper, the pinhole imaging technique[14,15] is applied to measure spatial position of a triple-pulse x-ray source produced by the Dragon-II LIA. The acquiring image denotes a two-dimensional distribution of the x-ray source, by which the spatial position of the source centroid can be obtained. Experiments are performed to observe spatial jitters of the source between different measurements. The interpulse differences of the source centroid are also measured in each experiment.

2. **Principle and setup**

In the experiment of flash radiography, photons emitted from the x-ray source pass through the objects placed in the light filed and finally reach the image receiving system for recording. The spatial distribution of the radiographic image $i(x,y)$ is the convolution of the spatial distribution of the source $s(x,y)$, the transmitted intensity distribution of the object $o(x,y)$ and the response of the detector to a point x-ray source $r(x,y)$, i.e.

$$i(x,y) = s(x,y) * o(x,y) * r(x,y), \qquad (1)$$

where the sign, $*$, denotes the convolution operation. The modulation transfer function (MTF) of the radiographic system is simply the product of the MTFs of each component[16], the relation of which can be obtained by making Fourier transform of Eq. (1) and given by

$$I(f_x, f_y) = S(f_x, f_y) \cdot O(f_x, f_y) \cdot R(f_x, f_y), \qquad (2)$$

where $I$, $S$, $O$ and $R$ stand for the MTFs of each corresponding component in Eq. (1). The sketch of pinhole imaging setup is illustrated in Fig. 1. A pinhole object is placed between the x-ray source and the image receiving system. For an ideal pinhole, the spatial intensity distribution $o(x,y)$ can be described by the delta function $\delta(x,y)$, which denotes the MTF of $O(f_x, f_y) = 1$. The geometrical magnification of the experimental arrangement $M$ is defined as the ratio of the pinhole-image distance $b$ to the source-pinhole distance $a$. If the blurring of image recording system is ignored in the setup with a large magnification, the relation of the source MTF and the image MTF will be simplified as

$$I(f_x, f_y) \approx S(f_x, f_y). \qquad (3)$$

It shows that the spatial distribution of the obtained image is exactly a reflection of the source. Then the relation between the source centroid $P_0(x,y)$ and the image centroid $P(x,y)$ can expressed as

$$P_0(x,y) = -P(x,y)/M, \qquad (4)$$

where the negative sign indicates opposite directions of the image position and the source position.

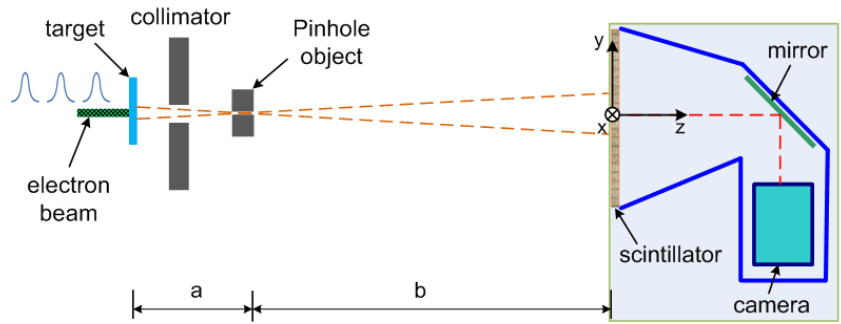

Fig. 1 Sketch of pinhole imaging setup.

## 3. Numerical simulations

The precondition of a tenable relation between the source and the image positions given by Eq. (4) is to image the source with an ideal pinhole, which is defined as an infinitesimally small hole through an infinitesimally thin and completely opaque sheet.[17] It cannot be realized in practice, especially when imaging penetrating radiation from the source. For MeV x-ray photons, high-Z materials are used to construct the pinhole object, which ought to be thick enough to make the "opaque" sheet. The pinhole aperture cannot be too small either because of the view field needed to cover enough area of the source. Besides, the influence of the screen blur should also be taken into consideration.

Here we perform numerical simulations to analyze the pinhole imaging process with the same or close parameters of the object and the experimental alignment. Firstly, the Monte Carlo method[18,19] is applied to simulate the generation of the x-ray source by striking an electron beam onto a target and the transmission of photons through the pinhole object. Then we calculate the convolution of the transmitted photon intensity distribution with the point-spread function (PSF) of the screen blur as the obtained image. The energy of electron beam is 19.0 MeV. The target of bremsstrahlung radiation is a 1.2-mm-thick tantalum. The pinhole object is made of a 65-mm-thick tungsten bar, in which the aperture of the hole through is 0.47 mm. The source-pinhole distance and the pinhole-image distance are $a = 1200$ mm, $b = 4600$ mm, respectively. The PSF of the screen blur is assumed to a Gaussian function with the standard deviation of 0.7 mm, the value of which is accordant with the measured result by the edge response technique[20].

In the simulation, a Gaussian function is used to model the spatial distribution of the source, the FWHM of which is firstly set to be 1.0 mm. The location of the source center (x, y) mm are set to be (2, 0), (-2, 0), (0, 2) and (0, -2), respectively. The images obtained by numerical simulations are shown in Fig. 2. Because the pinhole imaging process is not rotationally symmetric when the source is laterally decentered, we calculate the image centroid within different boundaries containing 90% and 50% of the photon intensity sum (PIS), respectively. The corresponding source centroid is worked out according to Eq. (4). Table 1 compares the calculated $P_0$ and the set $P_0$, between which the distance $\Delta L$ is superior to 0.1 mm with a 2-mm source displacement from the origin. The conditions of source FWHMs of 1.5 mm and 2.0 mm are then calculated and listed in the table. The results of calculated $P_0$ also show a good agreement with the results of set $P_0$.

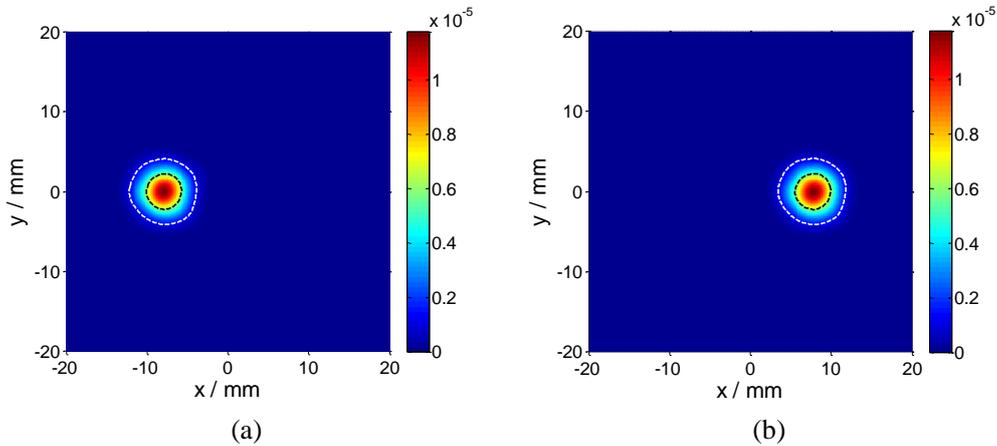

(a)   (b)

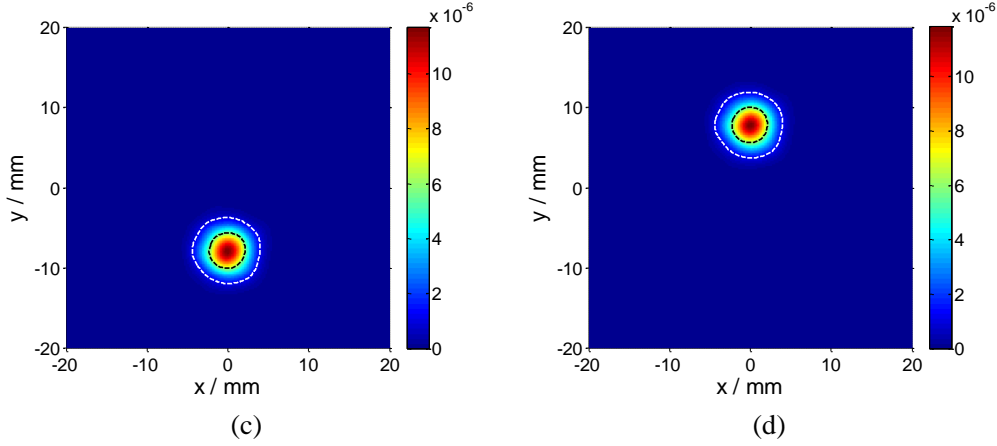

Fig. 2 Numerical simulations of pinhole images with different source positions. (a) x = 2 mm, y = 0 mm; (b) x = 2 mm, y = 0 mm; (c) x = 0 mm, y = 2 mm; (d) x = 0 mm, y = -2 mm. The white dash and the black dash denote the boundaries of 90% PIS and 50% PIS, respectively. The source FWHM is set to be 1.0 mm.

Table 1 Comparison of set and calculated source positions by numerical simulations. (unit mm)

| Source FWHM | Set $P_0(x,y)$ | Calculated $P_0(x,y)$ | | | |
|---|---|---|---|---|---|
| | | 90%PIS | ΔL | 50%PIS | ΔL |
| 1.0 | (2.000,0.000) | (2.063,0.002) | 0.063 | (2.056,0.002) | 0.056 |
| | (-2.000,0.000) | (-2.017,0.001) | 0.017 | (-2.024,0.003) | 0.024 |
| | (0.000,2.000) | (0.022,2.042) | 0.047 | (0.020,2.046) | 0.050 |
| | (0.000,-2.000) | (0.022,-2.039) | 0.044 | (0.014,-2.044) | 0.046 |
| 1.5 | (2.000,0.000) | (2.045,0.005) | 0.045 | (2.034,-0.002) | 0.034 |
| | (-2.000,0.000) | (-1.977,0.000) | 0.023 | (-1.989,0.005) | 0.012 |
| | (0.000,2.000) | (0.034,2.015) | 0.037 | (0.020,2.010) | 0.022 |
| | (0.000,-2.000) | (0.031,-2.011) | 0.033 | (0.018,-2.009) | 0.020 |
| 2.0 | (2.000,0.000) | (2.017,0.004) | 0.017 | (1.995,-0.011) | 0.012 |
| | (-2.000,0.000) | (-1.928,-0.002) | 0.072 | (-1.940,0.004) | 0.060 |
| | (0.000,2.000) | (0.044,1.978) | 0.049 | (0.019,1.967) | 0.038 |
| | (0.000,-2.000) | (0.040,-1.974) | 0.048 | (0.028,-1.972) | 0.040 |

## 4. Experimental measurements

Experiments are performed to observe the source position of the Dragon-II LIA which is able to generate triple x-ray pulses. A lead collimator is placed just in front of the radiographic source as a radiation shield of the area away from the central field. A tungsten bar with a pinhole through is precisely placed along the central axis (z-axis). The diameter of the pinhole is 0.47 mm and its thickness is 65 mm. The image receiving system consists of an LYSO scintillator screen, a flat mirror tilted at $45°$ with respect to z-direction, and a framing camera to record each pulse image. The parameters of the experimental alignment are $a = 1119$ mm, $b = 4681$ mm, which gives a magnification of $M = 4.183$. During the experiments, the energy of the electron beam is kept about 18.9 MeV and the current about 2.05 kA, trying to maintain a steady and identical state for all pulses.

Typical images of the triple-pulse x-ray source obtained by the pinhole imaging technique are shown in Fig. 3. In order to correct the pixel to pixel variations of the screen sensitivity and the dark current of the camera, the standard procedure for gain and offset modification is applied to the images.[21] Besides, relative displacements of the origin and azimuths of the axes are also corrected for different framing images. We calculate the source centroids of triple pulses considering both the image boundary of 90%PIS and that of 50%PIS. The experimental results of x-ray source centroid and parameter of the electron beam are listed in Table 2. For each pulse, the center and the radius of a minimum circle which contains all centroid positions in different measurements are used to denote the center and the range of source centroid jitters. For the boundary of 90%PIS, the jitter radii of the triple-pulse source centroid are $R_A = 0.155$ mm for pulse A, $R_B = 0.153$ mm for pulse B and $R_C = 0.072$ mm for pulse C, respectively. The deviation distances between the source-jitter centers of each two pulses are $L_{AB} = 0.690$ mm, $L_{BC} = 1.349$ mm and $L_{AC} = 0.674$ mm. For the boundary of 50%PIS, the jitter radii are $R_A = 0.188$ mm, $R_B = 0.213$ mm and $R_C = 0.125$ mm. And the deviation distances are $L_{AB} = 0.673$ mm, $L_{BC} = 1.295$ mm and $L_{AC} = 0.700$ mm. Experimental results show that the spatial jitters of the source centroid are relatively small for each pulse. But it is also seen that deviations of the source positions are distinct between different pulses, which most probably result from the corkscrew oscillation of the electron beam due to tilted beam injections, inaccurate alignments of solenoidal field as well as energy spread of electron beam.[22,23]

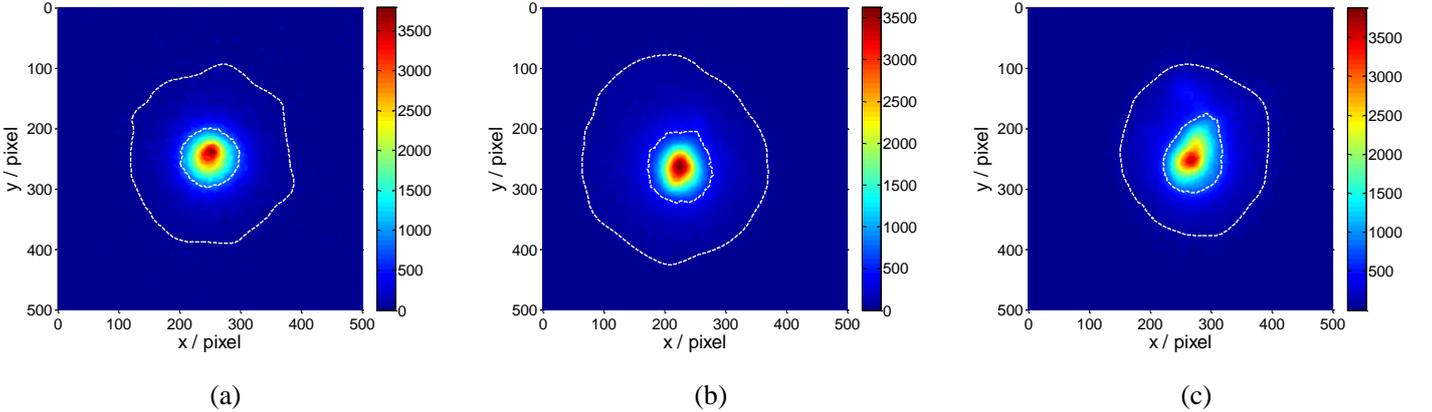

(a) (b) (c)

Fig. 3 Typical images of triple-pulse x-ray source by the pinhole method (No. #1). (a) Pulse A; (b) Pulse B; (c) Pulse C. The inside curve and the outside curve denote the boundaries of 50%PIS and 90%PIS, respectively.

Table 2 Experiment results of spatial position of the source centroid.

| Exp. No. | Pulse No. | E / MeV | I / kA | $P_0(x,y)$ / mm | |
|---|---|---|---|---|---|
| | | | | 90%PIS | 50%PIS |
| #1 | A | 18.9 | 2.04 | (0.000, -0.099) | ( 0.022, -0.073) |
| | B | 18.9 | 1.98 | (0.709, 0.230) | ( 0.627, 0.374) |
| | C | 18.8 | 2.07 | (-0.562, -0.296) | ( -0.532, -0.110) |
| #2 | A | 19.0 | 2.03 | (0.023, -0.341) | ( 0.007, -0.342) |
| | B | 18.8 | 2.04 | (0.685, 0.048) | ( 0.668, 0.195) |
| | C | 18.7 | 2.08 | (-0.567, -0.401) | ( -0.491, -0.261) |
| #3 | A | 19.0 | 2.05 | (-0.039, -0.353) | ( -0.037, -0.367) |

|  | B | 19.0 | 2.08 | (0.611, -0.032) | ( 0.561, 0.034) |
|  | C | 18.9 | 2.09 | (-0.633, -0.420) | ( -0.547, -0.254) |
|  | A | 19.0 | 2.06 | (0.041, -0.321) | ( 0.088, -0.347) |
| #4 | B | 18.9 | 2.06 | (0.659, 0.020) | ( 0.707, 0.148) |
|  | C | 18.9 | 2.06 | (-0.664, -0.387) | ( -0.658, -0.326) |
|  | A | 19.0 | 2.04 | (0.035, -0.354) | ( 0.044, -0.403) |
| #5 | B | 19.0 | 2.06 | (0.646, -0.071) | ( 0.659, -0.050) |
|  | C | 18.9 | 2.07 | (-0.603, -0.380) | ( -0.503, -0.236) |
|  | A | 19.0 | 2.05 | (0.150, -0.368) | ( 0.210, -0.399) |
| #6 | B | 19.0 | 2.08 | (0.730, -0.003) | ( 0.721, 0.110) |
|  | C | 19.0 | 2.07 | (-0.636, -0.398) | ( -0.536, -0.255) |

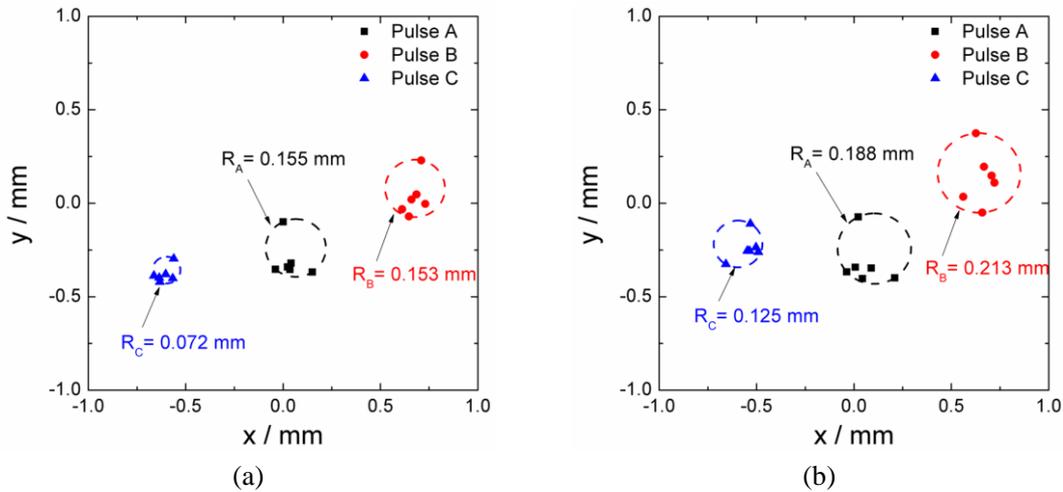

(a)          (b)

Fig. 4 Spatial jitters of triple-pulse x-ray source centroid. (a) Centroid within 90%PIS boundary; (b) centroid within 50%PIS boundary.

## 5. Conclusion

Spatial jitters of the triple-pulse x-ray source generated by the Dragon-II LIA are measured in the experiment based on the pinhole imaging technique. Numerical simulations are used to analyze the performance of the source position measurement with imperfect pinhole object and practical experiment alignment. Simulated results show a good agreement of the obtained source centroid with the set one. In each measurement, images of triple pulses are obtained, the centroids of which are calculated within 90%PIS and 50%PIS boundaries, respectively. Experimental results exhibit relative small spatial jitters of source centroid for each pulse whereas distinct centroid deviations between different pulses. For the application of large anti-scatter grid in the flash radiographic experiment, optimizations ought to be taken to provide an x-ray source with stable and uniform positions of all pulses, such as reduction of electron beam injection tilt, accurate alignment of the magnetic field and decrease in energy spread of electron beam.